\documentstyle{article}
\input epsf
\begin{document}
\parbox{13 cm}
{ \begin{flushleft} \vspace* {1.2 cm}
{\Large\bf {Engineering QND measurements for continuous 
variable quantum information processing}}\\
\vskip 1truecm {\large\bf { Matteo G A Paris } }\\ \vskip 5truemm
{Quantum Optics $\&$ Information Group, INFM Udr Pavia, Italy\\ 
{\tt E-mail:paris@unipv.it}, {\tt URL: www.qubit.it/\,$\tilde{}\,$paris}
} \end{flushleft} } \vskip 0.5truecm {\bf Abstract:\\} {
\noindent A novel scheme to realize the whole class
of quantum nondemolition (QND) measurements of a field quadrature is 
suggested. The setup requires linear optical components and squeezers,  
and allows optimal QND measurements of quadratures, 
which minimize the information gain versus state disturbance 
trade-off.} \vskip 0.1 cm 
\section{Introduction}\label{s:intro}
There is a growing interest for continuous variable (CV) quantum 
information processing, in particular for implementations based 
on manipulations of Gaussian states of light in optical 
circuits. Several quantum protocols \cite{furu,geza}, 
including teleportation, error 
correction, cloning and entanglement purification have been extended 
to CV systems, which may be easier to manipulate than quantum bits 
in order to accomplish the desired tasks \cite{plz}. \par  
In optical implementations, quantum information is encoded in values
of a single-mode field quadrature, say $x=1/2(a^\dag+a)$, $[a,a^\dag]$=1 being the 
mode operators. Therefore, in principle, the most relevant measurement 
for quantum protocols is provided by homodyne detection. However, 
the usual implementation of homodyning corresponds to a destructive 
detection, such that after the measurement we have no longer at disposal 
a quantum signal for further manipulations and/or measurements. 
It is thus of interest to devise a scheme for quantum nondemolition (QND)
measurements of a field quadrature, in particular for tunable QND 
measurements, in which the trade-off between information and disturbance may 
adjusted according to different needs. \par 
In this paper we suggest an all-optical scheme to realize the whole class
of QND measurements of a field quadrature \cite{xqnd}, from Von Neumann projective
measurement to fully non-demolitive, non-informative one. The setup involves
only linear optical components (including squeezers) and also allows an 
optimal QND measurement, which minimizes the information gain 
versus state disturbance trade-off. \par
The next section is devoted to the abstract description of a quantum
measurement, and to introduce two fidelities in order to quantify the state
disturbance and the information gain due to a measurement. In Section \ref{s:qnd} 
we analyze with some details the setup for QND measurements of a field 
quadrature. Section \ref{s:outro} closes the paper with some concluding remarks.
\section{Quantum measurements}\label{s:qmeas}
A generic quantum measurement is described by a set of {\em measurement
operators $\{M_k\}$}, with the condition $\sum_k M^\dag_k M_k=
{\rm I}$. The POVM of the measurement is given by $\{E_k\equiv M_k^\dag M_k\}$ 
whereas its quantum operation is expressed as $\varrho \rightarrow 
\sum_k M_k\varrho M_k^\dag$. This means that, if $\varrho$ is the initial 
quantum state 
of the system under investigation, the probability distribution of the outcomes 
is given by $p_k=\hbox{Tr}[\varrho\: E_k]= \hbox{Tr}[\varrho\: 
M^\dag_k M_k]$, whereas the conditional output state, after having
detected the outcome $k$, is expressed as $\sigma_k= \: M_k\varrho
M_k^\dag/p_k$, such that the overall quantum state after the measurement is
described by the density matrix $\sigma=\sum_k p_k \: \sigma_k=\sum_k 
M_k\varrho M_k^\dag$.  \par
Suppose you have a quantum system prepared in the state $\varrho$, and that 
you are interested in measuring the observable $K$. The so-called Von-Neumann 
(VN) measurement of $K$ is described by the operators $\{M_k=|k\rangle\langle 
k|\}$, where the $|k\rangle$'s are the eigenstates of $K$. Following the above 
prescription we have for VN measurements $p_k=\langle k|\varrho|k\rangle$, 
$\sigma_k=|k\rangle\langle k|$ and $\sigma= \sum_k p_k |k\rangle\langle k|$.
As a matter of fact, VN measurement (also called {\em projective} measurement) 
provides the maximum accessible information about the quantity $K$, at the price 
of {\em erasing} the quantum information of the state being investigated, which 
is no longer at disposal for further investigations or manipulations. 
\par
In opposition to projective measurement one may conceive a {\em nondemolitive}
measurement of $K$, which preserves the quantum state. This kind of measurement 
is described by the operators $\{M_k \propto I\}$, proportional to the identity 
operator. We have uniform $p_k$ and $\sigma=\varrho$, {\em i.e.} 
the quantum state is preserved, however the measurement is completely 
{\em non informative}. Overall, such kind of measurement may be viewed as a {\em 
blind} quantum repeater, which re-prepares any quantum state received at the 
input, without giving any information on its characteristics. 
\par
Between these two extrema there is a complete class of intermediate cases, 
{\em i.e.} quantum measurements providing only partial information about the
distribution $\{p_k\}$ while partially preserving the quantum state of the system. 
These schemes are sometimes referred to as QND 
measurements of the quantity $K$. 
\par
Let us now consider a generic quantum measurement $\{Q_k\}$ aimed 
to provide information about the quantity $K$. Two questions naturally arise 
about the characterization of its operation: \\
i) How much information is provided by the measurement ? Or, in other words, 
how close are the probability distributions $q_k=\hbox{Tr}[\varrho\: Q_k^\dag
Q_k]$ and $p_k=\langle k|\varrho |k\rangle$ ? In order to quantify this
resemblance we remind that the space of probability distributions 
$\{p_k\}_{k=1,...,M}$ is the M-simplex, where a privileged metric (the Fisher 
metric) exists and induces a distance between probabilities \cite{ks} given by 
$G=\left( \sum_k \sqrt{p_k\: q_k}\right)^2$ 
which represents a measure of the statistical distinguishability 
between the two distributions. \\ ii) How destructive is the measurement ? {\em I.e.} 
how far is the output state $\sigma$ to the input state $\varrho$ ? 
The {\em geometric} distance between two density matrices is given by 
$F=\left(\hbox{Tr}\left[\:\sqrt{\sqrt{\varrho}\:\sigma\sqrt{\varrho}}
\:\right]\right)^2$. This quantity is the proper generalization to mixed states 
of the standard quantum overlap used to quantify 
the (statistical, {\em i.e.} by measurements) distinguishability of pure states \cite{br}. 
F is also characterized by Uhlman theorem, which states that if $\varrho$ and
$\sigma$ are density matrices on a given Hilbert space ${\cal H}$, then
$F=\max_{\psi,\varphi}\: \left| \langle \psi | \varphi \rangle \right|^2$,
where $|\varphi\rangle$ and $|\psi\rangle$ are generic purification of
$\varrho$ and $\sigma$, {\em i.e.} pure states on ${\cal H} \otimes {\cal H}$
which have $\varrho$ and $\sigma$ as partial traces. If either $\varrho$ or 
$\sigma$ is pure F reduces to the standard overlap. For example, if 
$\varrho=| \Psi\rangle\langle\Psi|$ then we have 
$F=\langle\Psi|\sigma|\Psi\rangle$.
\section{QND measurements of a field quadrature}\label{s:qnd}
Our proposal to realize the whole class of QND measurements of a field
quadrature is depicted in Fig. 1. At first, the signal beam is mixed with a probe 
beam (which will be excited in squeezed vacuum state) in beam splitter of tunable
transmissivity $\tau_1$ ($BS_1$ in the figure). A tunable beam splitter 
can be easily implemented by a Mach-Zehnder interferometer. In the following we will write the  
transmittivity in the form $\tau_1=\cos^2\phi$. For the sake of simplicity, we consider
the measurements of the zero-phase quadrature $x=1/2(a^\dag+a)$; however, the same analysis is
valid for the generic $\theta$-quadrature 
$x_\theta=1/2(a^\dag e^{i\theta}+a e^{-i\theta})$.  \\ 
\begin{minipage}{6cm}
\begin{center}
\epsfxsize=6cm \epsfbox{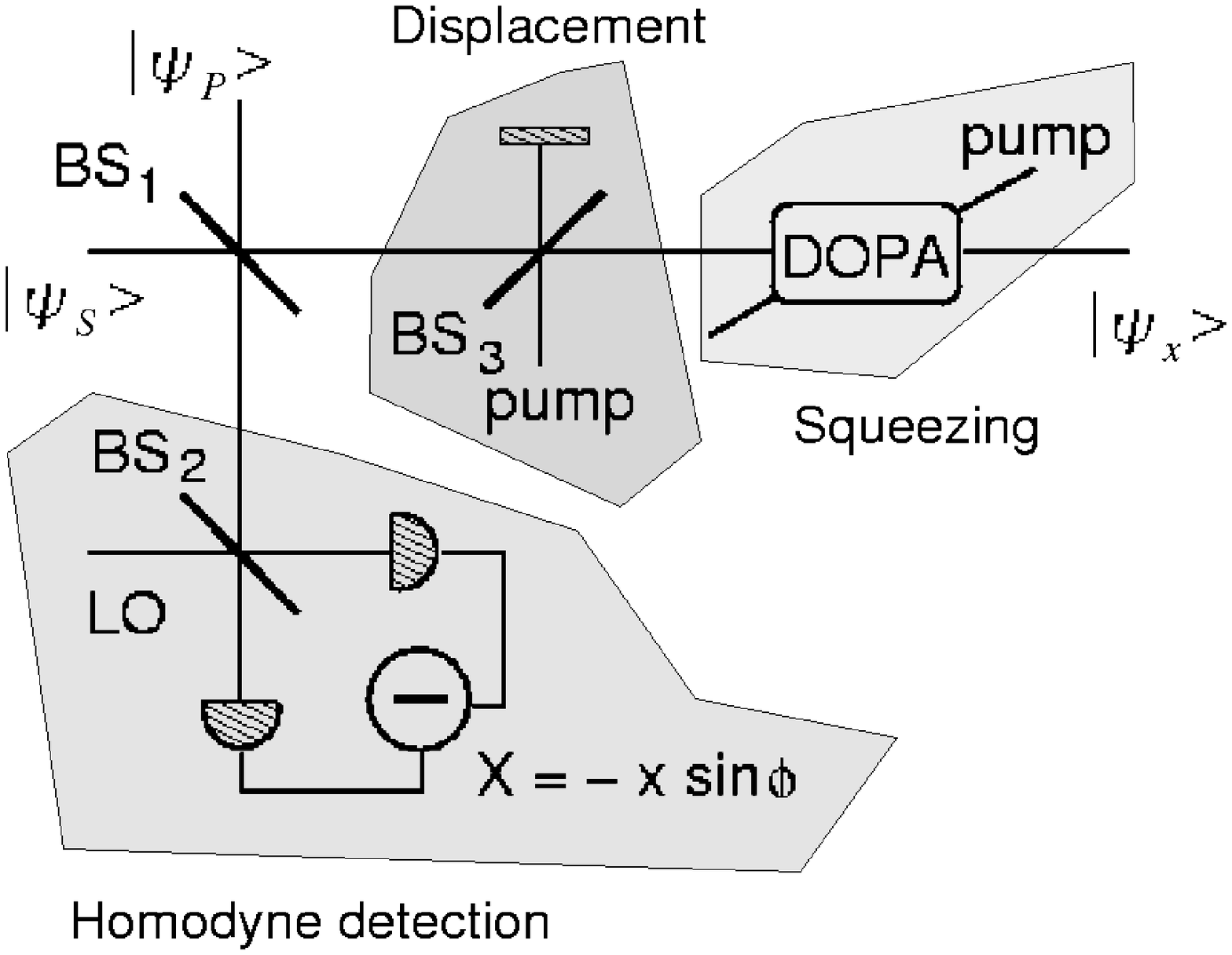}
\end{center} 
\end{minipage}
\begin{minipage}{8cm} {\small
{\bf Fig. 1}: Schematic diagram of the setup for QND measurements of a field
quadrature. The signal is mixed with a squeezed vacuum probe in a beam
splitter ($BS_1$) of tunable transmittivity $\tau_1$. The probe beam is then 
revealed by homodyne detection, whereas the signal beam is firstly displaced 
according to the value of the homodyne outcome, and then squeezed according to 
the transmittivity $\tau_1$ (see text). The local oscillator and the pumps
for the displacing and squeezing stages are provided by a common laser
source.}
\end{minipage} \\ $ $ \\
After the beam splitter, the probe beam is revealed by homodyne detection.
Taking into account the reflectivity amplitude of $BS_1$, from an outcome 
$X$ from the homodyne we infer a value $x=-X/\sin\phi$ for the quadrature 
of the signal beam. The signal is then displaced by an amount
$\alpha^*=-X\tan\phi=x\tan\phi\sin\phi$, by feedback of the outcome of the
homodyne detector, and finally squeezed by an amount $r^*$ such that 
$\exp\{r^*\}=\sqrt{\tau_1}=\cos\phi$.
The displacement transformation $D(\alpha^*)=\exp\{\alpha^* a^\dag -
\bar\alpha^* a\}$ can be obtained by mixing the signal with a strong pump beam 
excited in strong coherent state $|z\rangle$ $|z|\rightarrow\infty$ (which 
may be by the local oscillator of the homodyne detector) in a 
beam splitter ($BS_3$ in the figure) with transmittivity approaching unit 
value $\tau_3 \rightarrow 1$ such that $\alpha^*=z\sqrt{1-\tau_3}$.
The squeezing transformation is obtained with a degenerate optical parametric
amplifier (DOPA), where the pump mode is again provided by the common laser source 
providing the homodyne LO. Since the cosine is smaller than one, $r^*$ is negative, 
and this means that we are squeezing the signal in a direction orthogonal to 
the quadrature we are going to measure. \par
The probability density for the inferred values of the signal 
quadrature is given by 
\begin{eqnarray}
p(x)=-\sin\phi \: q(X) = \tan\phi \int dy\: |\psi_{\sc s}(y)|^2 \: 
\left|\psi_{\sc p}\left[\tan\phi(y-x_0)\right] \right|^2
\label{probx0}\;,
\end{eqnarray}
where $q(X)=\hbox{Tr}[|\psi_{\sc s},\psi_{\sc p}\rangle
\rangle\langle\langle\psi_{\sc p},\psi_{\sc s}|\:
I \otimes |X\rangle\langle X|]$ is the probability density for the homodyne outcomes, and
$\psi_j(x), \: j={\sc s,p}$ are the signal and probe wave-functions in the
quadrature representation, {\em i.e.} $|\psi_j\rangle=\int dx\; \psi_j(x)\: |x\rangle$.
The conditional output state, after having inferred the value 
$x$ for the signal quadrature, is given by 
\begin{eqnarray}
|\psi_x\rangle = \sqrt{\frac{\sin\phi}{p(x)}}\: S(r^*)\: D(\alpha^*)\: \langle
-x \sin\phi | V_\phi |\psi_{\sc s},\psi_{\sc p}\rangle\rangle 
\label{condx}\;,
\end{eqnarray}
where $V_\phi=\exp\{i\phi(a^\dag b+ b^\dag a)\}$ is the evolution operator of
the beam splitter $BS_1$. 
For a probe mode excited in a squeezed vacuum we
have $\psi_{\sc p}(x)=(2\pi\Sigma^2)^{-1/4}\exp \{-\frac{x^2}{4\sigma^2_{\sc p}}\}$, 
where the variance is given by $\sigma_{\sc p}^2=1/4 \exp\{\pm 2r\}$ according to the 
direction of squeezing. We refer to as a {\em squeezed} probe for the minus sign
(squeezing in the direction of the quadrature to be measured) and to as an {\em
antisqueezed} probe for the plus sign (squeezing in the orthogonal direction).
The average number of photons carried by the probe is given by $N_{\sc
p}=\sinh^2 r$ in either cases. After minor algebra we get 
\begin{eqnarray}
p(x)&=&|\psi_{\sc s}(y)|^2 \star G(y,x,\sigma^2_{\sc p}/\tan^2\phi) \\
|\psi_x(y)|^2&=&\frac{1}{p(x)}\:|\psi_{\sc s}(y)|^2\:G(y,x,\sigma^2_{\sc p}/\tan^2\phi)
\label{sqcond}\;,
\end{eqnarray}
where $\star$ denotes convolution and $G(y;x,\sigma^2)$ a Gaussian of 
mean $x$ and variance $\sigma^2$. \par
For $\sigma^2_{\sc p}/\tan^2\phi \rightarrow 0$ {\em i.e.} either for strongly
squeezed probe or for an almost transparent $BS_1$ we have $p(x)\rightarrow 
|\psi_{\sc s}(x)|^2$ and $|\psi_x (y)|^2 \rightarrow \delta(y-x)$, which means
that we are approaching a projective VN measurement of the quadrature. 
On the other hand,  for $\sigma^2_{\sc p}/\tan^2\phi \rightarrow \infty$ (strongly
antisqueezed probe or an almost opaque $BS_1$) we may write $p(x)$ as 
a very broad Gaussian, and we have $|\psi_x (y)|^2 \rightarrow |\psi_{\sc s}(y)|^2$,
that is we are approaching a non-informative blind quantum repeater.
By tuning either the probe squeezing parameter or the transmittivity of $BS_1$ 
we may also realize the whole class of intermediate QND measurements of the quadrature.
\par
For Gaussian signals the two fidelities $F$ and $G$, which measure that state
disturbance and the information gain respectively, may be easily evaluated in
terms of the single variable $x=\sigma_{\sc p}/(\sigma_{\sc s}\tan\phi)$, 
$\sigma_{\sc s}^2$ being the variance of the signal' wave-function.
We have $$ F=\frac{\sqrt{2}x}{\sqrt{1+2x^2}} \quad \hbox{and} \quad 
G=2\:\frac{\sqrt{1+x^2}}{2+x^2}\:.$$
Of course we have $F\rightarrow 0$ and $G\rightarrow 1$ for $x \rightarrow 0$,
and vice-versa for $x\rightarrow \infty$. However, in general 
the quantity $F+G$ is not constant, and this means that by
varying the squeezing of the probe we obtain  different trade-off between information 
gain and state disturbance. An optimal choice of the probe, corresponding to 
maximum information and minimum disturbance, maximizes $F+G$. 
The maximum is achieved for $x\equiv x_{\sc m}\simeq 1.2$, corresponding 
to fidelities $F[x_{\sc m}]\simeq 86 \%$ and $G[x_{\sc m}]\simeq 91\%$.
Notice that for a chosen signal, the optimization of the QND measurement can 
be achieved by tuning the internal phase-shift of the interferometer, without the 
need of varying the squeezing of the probe. For a nearly balanced interferometer we have 
$\tan\phi\simeq 1$: in this case the optimal choice for the probe is a state slightly 
anti-squeezed with respect to the signal, {\em i.e.} $\sigma_{\sc p} \simeq 1.2\: 
\sigma_{\sc s}$. Finally, the fidelities are equal for $x\equiv x_{\sc e}\simeq 1.3$,
corresponding to $F[x_{\sc e}]=G[x_{\sc e}]\simeq 88 \%$.
For non Gaussian signals the behavior is similar though no simple analytical
form can be obtained for the fidelities. In this case, in order to find the 
optimal QND measurement, one should resort to numerical means. 
\section{Conclusions}\label{s:outro}
In conclusions, we have suggested a novel scheme assisted
by squeezing and linear feedback to realize an arbitrary QND 
measurement of a field quadrature. Compared to previous QND 
proposals \cite{pre} the 
main features of our setup can be summarized as follows: i) it involves 
only linear coupling between signal and probe, ii) only single mode 
transformations on the conditional output are needed.
\par
The present setup permits, in principle, to achieve both a projective and a
fully non-destructive quantum measurement of a field quadrature.  In practice,
however, the physical constraints on the maximum amount of energy that can be
impinged into the optical channels pose limitations to the precision of the
measurements. This agrees with the facts that both an exact repeatable
measurement and a perfect state preparation cannot be realized for observables
with continuous spectrum \cite{oza}. 
\par 
Compared to a vacuum probe, the squeezed/anti-squeezed 
meters suggested in this paper provide a consistent noise reduction in the desired
fidelity figure already for moderate input probe energy. In addition, by
varying the squeezing of the probe an optimal QND measure can be achieved,
which provides the maximum information about the quadrature distribution of
the signal, while keeping the conditional output state as close as possible to
the incoming signal.  
\par
In order to tune the setup and achieve the whole 
class of QND measurements we have two independent parameters at disposal: 
the probe squeezing parameter and the transmittivity of $BS_1$. This is a
another relevant feature of the scheme, since a too large squeezing would increase too
much the energy impinged into the apparatus, whereas a tuning based only on
the transmittivity would largely affect the detection rate.
\section*{Acknowledgments} This work has been cosponsored by the INFM through 
the project PRA-2002-CLON, and by EEC through the TMR project IST-2000-29681 (ATESIT). 

\end{document}